\documentclass{llncs}

\usepackage[utf8]{inputenc}
\usepackage{microtype}
\usepackage{textcomp}
\usepackage{todonotes}
\usepackage[british]{babel}
\usepackage{hyphenat}

\usepackage[hyphens]{url}

\expandafter\def\expandafter\UrlBreaks\expandafter{\UrlBreaks%  save the current one
  \do\a\do\b\do\c\do\d\do\e\do\f\do\g\do\h\do\i\do\j%
  \do\k\do\l\do\m\do\n\do\o\do\p\do\q\do\r\do\s\do\t%
  \do\u\do\v\do\w\do\x\do\y\do\z\do\A\do\B\do\C\do\D%
  \do\E\do\F\do\G\do\H\do\I\do\J\do\K\do\L\do\M\do\N%
  \do\O\do\P\do\Q\do\R\do\S\do\T\do\U\do\V\do\W\do\X%
  \do\Y\do\Z}

\newcommand{\safescrum}{SafeScrum\textsuperscript{\tiny\textregistered}}
\newcommand{\rscrum}{R-Scrum}
\newcommand{\rqone}{Which common principles and practices can be derived from existing approaches for agile development of safety-critical systems?}
\newcommand{\rqtwo}{Which practical challenges exist when applying these principles and practices in a large-scale industrial setting?}

\usepackage[normalem]{ulem} % for \sout
\usepackage{xcolor}

\usepackage{ifthen}
\usepackage{amssymb}
\newboolean{showcomments}
\setboolean{showcomments}{true} % toggle to show or hide comments
\ifthenelse{\boolean{showcomments}}
  {\newcommand{\nb}[2]{
    \fcolorbox{gray}{yellow}{\bfseries\sffamily\scriptsize#1}
    {\sf\small$\blacktriangleright$\textit{#2}$\blacktriangleleft$}
   }
   
  }
  {\newcommand{\nb}[2]{}
   
  }

\begin{document}
\mainmatter              % start of a contribution
\title{Challenges of Scaled Agile\newline{} for Safety-Critical Systems}
\author{Jan-Philipp Steghöfer\orcidID{0000-0003-1694-0972} \and Eric Knauss\orcidID{0000-0002-6631-872X} \and Jennifer Horkoff\orcidID{0000-0002-2019-5277} \and Rebekka Wohlrab\orcidID{0000-0002-5449-7900}}
\authorrunning{Steghöfer et al.} % abbreviated author list (for running head)
\institute{Chalmers $|$ University of Gothenburg, Gothenburg, Sweden\\
Software Engineering Division, Department of Computer Science and Engineering\\
%\email{jan-philipp.steghofer@gu.se}, \email{eric.knauss@gu.se}, \email{jenho@chalmers.se}, \email{wohlrab@chalmers.se}
\email{\{jan-philipp.steghofer, eric.knauss\}@gu.se}, \email{\{jenho, wohlrab\}@chalmers.se}}

\maketitle              % typeset the title of the contribution

\begin{abstract}
%\textbf{[Context]} 
Automotive companies increasingly adopt scaled agile methods to allow them to deal with their organisational and product complexity. Suitable methods are needed to ensure safety when developing automotive systems. On a small scale, \rscrum{} and \safescrum{} are two concrete suggestions for how to develop safety-critical systems using agile methods. However, for large-scale environments, existing frameworks like SAFe or LeSS do not support the development of safety-critical systems out of the box.
%\textbf{[Objective]}
We, therefore, aim to understand which challenges exist when developing safety-critical systems within large-scale agile industrial settings, in particular in the automotive domain.
%\textbf{[Method]}
Based on an analysis of \rscrum{} and \safescrum{}, we conducted a focus group with three experts from industry to collect challenges in their daily work.
%\textbf{[Results]}
We found challenges in the areas of living traceability, continuous compliance, and organisational flexibility. Among others, organisations struggle with defining a suitable traceability strategy, performing incremental safety analysis, and with integrating safety practices into their scaled way of working.
%\textbf{[Conclusion]}
Our results indicate a need to provide practical approaches to integrate safety work into large-scale agile development and point towards possible solutions, e.g., modular safety cases.

 % large-scale agile as another keyword?
\end{abstract}
\section{Introduction}

In the automotive domain, several dozen development teams work together in a highly coordinated fashion towards the delivery of a product.
Systems and software engineering need to be combined in these cases to deliver a final product and the chosen process needs to scale across a large number of teams and different engineering disciplines.
To manage this complexity, many companies have started adopting solutions such as the Scaled Agile Framework (SAFe)\footnote{\url{https://www.scaledagileframework.com/}} or Large-Scale Scrum (LeSS)\footnote{\url{https://less.works/less/framework/index.html}}. These agile frameworks also aim at reducing the time-to-market and propose solutions for the coordination between teams, exchange of artefacts, and the prioritisation of work within an organisation.

However, these frameworks do not provide explicit support for the creation of safety-critical systems and the risk management, safety analysis, and certification activities associated with ensuring safety. 
On a small scale, \rscrum~\cite{rscrum} and \safescrum~\cite{hanssen2018safescrum} help organisations to combine the documentation needs and the rigour required for safety-critical systems with agile development.
However, the existing approaches do not describe how to scale them beyond individual teams.
It is also not obvious how to tack on such activities, in particular since safety is a cross-cutting concern which needs to be addressed on all levels of the system. Preparing a release for certification in a big-bang approach, i.e., after development has mostly finished, has also proven to be infeasible. Therefore, safety should be considered continuously and integrated in the everyday work of the engineers to produce all necessary artefacts and the required audit trail in an ongoing fashion. 
Understanding how to do this in a practical setting is an important first step towards extending agile frameworks like SAFe and LeSS.

In parallel with the advent of methods to scale agile, variations of agile approaches that address safety-critical systems have been developed. Two well-cited examples are \rscrum~\cite{rscrum} and \safescrum~\cite{hanssen2018safescrum}, but other, less comprehensive approaches exist. Kasauli et al.~\cite{kasauli2018safety} provide an overview of the literature in this area and identify a number of solution approaches to combine agile practices, mostly from Scrum and XP, with activities necessary for safety-critical systems such as risk management and hazard analysis. Other authors are addressing specific domains (e.g., medical devices~\cite{schoonderwoert2018agile}) to provide experiences from practice.

However, a number of challenges remain in adapting the processes of large organisations producing safety-critical systems to fulfil both the need for agility and the required rigour for certification. This is a particular problem in the automotive domain since the organisations involved in producing vehicles are very large, distributed over many disciplines and physical locations, and have established practices and tool-chains that are difficult to change~\cite{Broy2007,Ebert2017}. This leads us to the following research questions:

\begin{description}
\item[RQ1:] \rqone
\item[RQ2:] \rqtwo
\end{description}

%\rebekka{large-scale agile or scaled-agile? Or large-scale companies using agile approaches? (well, some even use hybrid approaches with some plan-driven elements?)}
%\jenho{the RQs say nothing about scale, not so consistent with intro. Maybe: large-scale industrial setting.}
To answer RQ1, we analysed existing literature with a focus on \rscrum{} and \safescrum{}, as well as the overview presented in Kasauli et al. \cite{kasauli2018safety}. 
We derived the common principles \emph{focus on traceability}, \emph{safety as an ongoing set of activities}, \emph{shared responsibility of the team}, as well as \emph{involvement of assessors or auditors in ongoing development} (see Section~\ref{sec:agile-approaches}). 
We also identified that existing approaches \emph{do not address scaling beyond a single team}, \emph{have no provisions for systems with mixed criticality}, and \emph{lack concrete approaches for automation}. %\jenho{forward reference to section with full results?  It's strange to give some results here, makes me wonder if this is it.}

Based on these findings, we conducted a focus group to answer RQ2. We presented the common principles to practitioners in the domain and elicited a number of challenges that occur in the automotive industry in practice when agile methods are used to create safety-critical systems. 
These challenges can be mapped to the areas of \emph{living traceability}, \emph{continuous compliance}, and \emph{organisational flexibility} (see Section~\ref{sec:open-challenges}). There, we also compare the challenges to what is known from the literature.% \jenho{present challenges in Section X.  Here in this paper, not right here.}

Thus, the main contribution of this paper is an overview of the unsolved, practical challenges in combining agile methods with safety-critical systems with a particular focus on large-scale development efforts in the automotive domain. 
This will create the foundation for future work in which specific solution approaches to these challenges will be developed and evaluated.

%\todo[inline]{A skeptical person (e.g., from RE) would ask why safety critical development is trying to be agile anyway?  Do we need to justify that agility is desired even for saftey-critical systems?.}
%\todo[inline]{Reading this, I expect us to do more with SAFe and LeSS.  It seems logical to use our findings to extend/fix SAFE/LeSS, but we don't do this.  Need to emphasize that although this may be the ultimate goal, there is an important first step of understanding needs in practice, this is what we do here.}

%\section{Background and Related Work}

%From Francis:  “agile methods for safety-critical systems”  Van schooenderwoert, Shoemaker.  AAMI- guidance on agile in healthcare.

\section{Methodology}
\label{sec:methodology}
%\jenho{the two paragraphs above already describe some of the methods, so this reads as repetitive.  Also, if we really want to describe methodology, we should describe the extraction of principles, not just the workshop.  I may suggested to take the two paragraphs above and merge them into this section.}

We base our findings in this paper on a focus group in which we discussed \rscrum{} and \safescrum{} with practioners. % with practitioners from two large, international companies from the automotive domain. In addition, a practitioner from a large international supplier of medical devices was present. %\jenho{two large, international companies?}
To prepare this focus group, one of the researchers extracted a list of commonalities and differences from the published accounts of \rscrum{} and \safescrum{} (e.g., \cite{rscrum,hanssen2018safescrum}). This list was discussed and refined with the other co-authors and contrasted with the findings of Kasauli et al.~\cite{kasauli2018safety}. It was decided to focus on the commonalities of the approaches since they show the common underpinnings of how agile approaches can be applied to safety-critical products. These results provide the answer to \textbf{RQ1} and are presented in Section~\ref{sec:agile-approaches}. %\jenho{ah, here it is, just in a different order than expected, could be OK}

%\jp{I am leaving out the mapping to the other workshop here since I feel it's going to be too much to describe this here as well. What do you think?\eric{I feel like it will be important to keep this a little bit disjoint to the extension of the SEAA paper. Thus, less is likely better.}}

The focus group included three industrial experts, all with several years of experience in developing safety-critical systems in agile settings. These experts are leaders in the field and are all involved in strategic projects transforming their respective international organizations towards agile development. Two of the practitioners are experts on process, methods, and tools from two different automotive OEMs. One of them has a background in safety assurance and the other in scaled agile and AUTOSAR-compliant tool chains. The third practitioner is a senior software engineer in a medical device company with a background in agile in industrial, regulated environments. Although from a different domain, this expert participated in the workshop due to their experience and strong interest in scaled safety-critical agile methods. Including an expert from another domain also allowed pinpointing which challenges are specific to automotive. 

As part of the focus group, we presented \rscrum{}, \safescrum{}, and the findings from~\cite{kasauli2018safety} to the practitioners and gauged their reaction to them. Based on this presentation, we invited the practitioners to share the challenges they see with applying agile methods to safety-critical systems within the large-scale development efforts in their organisations. The practitioners then brainstormed and described their own experience and the challenges they encounter. The researchers took extensive notes and collected the remarks on post-it notes. When saturation was achieved, these post-it notes were roughly sorted into topical areas by one of the researchers. These areas were then discussed in the group and topics were moved between topical areas when necessary to jointly create a clustering, thereby excluding or merging topics that were closely related.

After the conclusion of the focus group, the researchers ensured that agreement was reached on all topics when they reconstructed the discussion and recorded the findings based on their notes as well as the final clustering. 
They are presented in Section~\ref{sec:open-challenges}. 
%The researchers ensured that agreement was reached on all topics. 
These results provide the answer to \textbf{RQ2} and are presented in Section~\ref{sec:open-challenges}.
All results were member checked with the participants of the original workshop, %in a follow-up presentation. They 
who corrected some details but confirmed the overall findings.

%\jp{Hopefully, we will be able to do some member checking}
%\rebekka{can we do a quick check by phone/email with the three focus group people?}
%Member checking
%something about the data you got first from the literature, as it's presented below

\section{Existing Agile Approaches for Safety-Critical Systems}
\label{sec:agile-approaches}

There are two agile approaches that cover the entire development lifecycle for safety-critical systems in the literature: \emph{\rscrum}~\cite{rscrum} and \emph{\safescrum}~\cite{hanssen2018safescrum}.

\rscrum{} is described by Brian Fitzgerald and colleagues from their observations at the company QUMAS which builds ``compliance management solutions''. The paper thus does not describe a method that has been designed by researchers and evaluated in a company, but is rather a collection of the best practices at QUMAS that have proven worthwhile over the years. There are no studies to validate the usefulness of \rscrum{} outside of QUMAS.

\safescrum{} in turn is a more designed process in which researchers and practitioners created a version of Scrum to fit the needs of safety-critical projects with a focus on IEC 61508:2010. The process (or rather ``method framework'' as the authors call it) has been used in a number of case studies. It is well-documented in a book~\cite{hanssen2018safescrum} and a number of case studies (e.g., \cite{staalhane2012application,myklebust2016agile}) that demonstrate its application and efficacy.

In addition to this, Kasauli et al.~\cite{kasauli2018safety} describe the results of a systematic mapping study and present an exhaustive list of relevant research about applying agile methods to safety-critical systems. The authors also provide challenges and solution candidates that have been reported in the literature which have been validated in a workshop with industrial practitioners. These solution candidates are, however, not embedded in a process or method framework.

A comparison of these sources shows that the proposed solutions for using agile methods to develop safety-critical systems share a number of commonalities:
\begin{description}
	\item[Focus on traceability:] Traceability is regarded as a foundation for the ability to certify software. \rscrum{} makes \emph{living traceability} a cornerstone of the method to provide ``complete transparency into the development process at any point in time''~\cite{rscrum}. \safescrum{} also emphasises traceability, in particular to fulfil the requirements of the IEC 61508:2010 standard, but also to enable change impact analysis and to perform safety testing~\cite{hanssen2018safescrum}. Kasauli et al.~\cite{kasauli2018safety} also report on two sources explicitly stating the need for traceability to ensure requirements are met and to determine which tests need to be run. 
	\item[Safety as an ongoing set of activities:] In order to ensure that safety is taken into consideration in all design decisions and in the daily programming and validation work, it is integrated into the process tightly and activities involving safety are performed continuously rather than at discrete points in time (e.g., immediately before a release). \rscrum{} aims to achieve \emph{continuous compliance} by including risk analysis in user story prioritisation and including a quality control board that is involved in continuously checking the developed code as well as accompanying documentation and design documents. In addition, quality assurance audits are included in each sprint and additional ``hardening sprints'' can be scheduled before releases. \safescrum{} likewise introduces safety into the sprint planning and the sprints themselves. An ``alongside engineering team'' is responsible for these activities that include updating the hazard log and safety cases, performing risk analysis and safety validation, and ensuring that safety requirements are captured. Kasauli et al.~also mention suggestions from the literature to include safety considerations in Scrum ceremonies such as daily stand-up meetings, setting up a continuous integration tool-chain that includes safety builds, including relevant documents such as hazard logs in code reviews, and perform continuous risk management.
	\item[Shared responsibility of the team:] Notably, all three sources suggest that the development team itself is involved in the activities to ensure safety, not a separate group of people. Even the ``alongside engineering team'' from \safescrum{} should not be seen as a team separate from the developers, but rather defines roles that can be fulfilled by the developers themselves. The book states, however, that this ``may involve others external to the \safescrum team''. In any case, developers are never absolved from taking responsibility for safety in the ongoing safety activities. They need to be able to work with the risk analysis, update hazard logs and other artefacts, and ensure in their design decisions that safety considerations are upheld. They are also responsible for writing appropriate test cases for safety validation. Kasauli et al.\ explicitly list literature that mentions collective code ownership, experts in the team, and the necessity that team members are familiar with safety standards in addition to the joint activities mentioned above.
	\item[Involvement of assessors or auditors in ongoing development:] While not taken up by Kasauli et al., both \rscrum{} and \safescrum{} suggest to include assessors or auditors for the final product in the development process. In the case of \rscrum{} at QUMAS, these audits are performed by the customers and include the development process itself. This means that the organisation ensures that their process adheres to the standards set by the customers which usually follow the established safety standards in turn. In \safescrum, repeated safety audits are called for to ensure independent validation of the created product and process. In case of both methods, an established, traceable audit trail facilitates these occasions greatly.\sloppy
\end{description}

However, there is \emph{no notion of scaling beyond a single team}. Neither \rscrum{} nor \safescrum{} provide guidance how work on safety should be divided between collaborating teams or between a product and team level. They locate the responsibility for safety of the product with the single development team (and the auxiliary ``alongside engineering team''), but also assume that this team can control the entire development lifecycle of the product. This is insufficient in situations in which a complex organisational structure is used in which safety has to be ensured across large number of teams working on the same product. While a diagram describing the activities of the alongside engineering team in \safescrum{} does contain the item ``subcontractor'' management, this issue is not taken up in the rest of the book~\cite{hanssen2018safescrum}. Likewise \rscrum{} and the descriptions by Kasauli et al.\ lack details of how to involve suppliers apart from including external actors in planning and review meetings.

Furthermore, there is \emph{no notion of mixed criticality}. Both \rscrum{} and \safescrum{} assume that the product in its entirety is safety-critical and that all parts of the product thus need to be treated as safety-critical. In reality, however, products often consist of particular, safety-critical parts that are combined with other, non-critical components. Applying the same process to both kinds of assets can result in additional cost since the overhead necessary to ensure that the safety-critical parts can be certified is unnecessary for the non-safety-critical ones.

Finally, there are \emph{no guidelines on the automation of safety certification}. In practical settings, tool support is required to ensure that activities concerned with safety can be embedded in the development process. This is particularly true for complex software product lines, e.g., in the automotive industry: all variants of a highly variable product need to be safe. Thus, safety cases need to be applicable to all variants and can become shared assets or even contain variability information themselves. Such scenarios require tool support and automation.

In summary, we extracted the following remaining issues that need to be addressed in mature domains such as automotive from our analysis of \rscrum{} and \safescrum{} as well as the partial solutions reported by Kasauli et al.:
\begin{description}
		\item[Scaling safe Scrum:] combining the scalability of SAFe or LeSS with the safety features of \rscrum{} or \safescrum{} for multi-team projects;
		\item[Mixed criticality:] safety-critical parts of products need to be developed with more ceremony than parts that are not safety-critical;
		\item[Automation:] automate generation of ``proof of compliance'' documentation within complex Continuous Integration/Deployment (CI/CD) tool-chains.
	\end{description}

\section{Open Challenges According to Industry}
\label{sec:open-challenges}

Upon presenting and discussing the principles and practices of currently safety-focused agile methods in our focus group, the focus group members brainstormed the challenges they encounter in their organisations. We categorised these challenges into three different areas that need to be addressed for scaled agile for safety-critical systems to become a reality in industry. The first two of these areas overlap with the solution areas of current frameworks listed in Section~\ref{sec:agile-approaches}. However, we describe specific and detailed challenges for those and take an additional step by introducing challenges on the organisational level.

\begin{description}
	\item[The foundation: living traceability.] As recognised in both \rscrum{} and \safescrum, traceability\,---\,and, in particular, the ``living'' version of it\,---\,is the foundation for an agile way of working with safety. The ability to connect the individual artefacts in the development process to each other enables the generation of the reports required by safety standards and facilitates the construction of safety cases. This goes beyond the traceability between requirements and test cases prescribed by safety standards, though: living traceability means that developers actively and continuously create, maintain, and delete trace links while they go about their development work. The resulting network of trace links not only supports safety, it also helps the developers with change impact analysis, program comprehension, and identifying technical debt. % \rebekka{keep the summary even shorter here and move details, quotes etc.~to the sections below?} % JH: I don't understand the second half of this quote.  Safety motivates all of our needs?  JP: I understand it as ``Safety is the one thing where all of our traceability needs become relevant''
	\item[The goal: continuous compliance.] The goal for all organisations that have been a part of this study as well as for those \rscrum{} and \safescrum{} have been applied to is to continuously produce the necessary safety arguments to ensure that compliance can be proven at any point in the development process. This is in contrast to the established way of working where the safety arguments are produced in a big bang approach towards the end of the development cycle or even immediately before an audit or certification. Continuous compliance enables an organisation to show at any point in time that their system complies to all necessary standards and has been developed following a process able to produce a safe system.
	%JH: may want to say something about:  The goal here is to produce arguments over the deltas of change, only focusing on changes and their impacts.  Open questions include: how often to create a new safety argument?  How big must the delta be before safety must be argued?  How to balance between a desire for continual safety and the resources needed to produce frequent safety arguments?  How often is often enough? JP: Used below.
	\item[The next step: organisational flexibility.] Once continuous compliance is achieved, the final stepping stone is to achieve flexibility in the organisation to work within a safety-critical domain in a truly agile way. This flexibility has to be achieved in three different areas: the \emph{ecosystems} of components that are being used and exchanged with suppliers, \emph{change management} within the organisation, and the \emph{way of working} with critical artefacts.  %JH: for me the last two are two vague, change management is counter to flexibilty?  And the way of workind could or could not effect flexiblity. JP: reformulated this to make it clearer (hopefully)
\end{description}

%In the following, we will describe these challenges in more detail. 

\subsection{Living Traceability}

Continuously maintained traceability provides the foundation for scaled agile for safety-critical systems. As one of the workshop participants put it: ``There are many motivators for traceability, but safety captures all of our needs'', meaning that all needs for traceability from other areas of development are also present when discussing the needs for traceability to ensure safety. Our participants identified the following challenges in this area:

\paragraph{``Select the right direction for traceability.''} Establishing a traceability information model (TIM) that supports safety analysis is a challenge. More fine-grained artefacts (lower-level artefacts) should contain links that link to more abstract artefacts (higher-level artefacts). One reason for this is variability: an abstract, high-level artefact can be refined into several variants on the lower level. Trace links from the high-level artefact to all variants are impractical, so instead, each variant should link to the higher-level artefact. That makes tool-support to collect the links crucial and needs to be captured in the TIM which defines the structure and semantics of the trace links. A common misunderstanding of bidirectional traceability is that trace links must exist in both directions\,---\,instead, tools must exist that can reconstruct one direction from the other if necessary. 

\paragraph{``Provide a meaningful TIM for safety-critical systems.''} Defining a traceability information model that supports the required semantics for safety-critical systems can pose a problem (see also next item). A suitable TIM needs to connect all safety-related artefacts, such as requirements, safety cases, and tests, in order to detect inconsistencies between them, to allow tracking their evolution, and to show that all safety concerns have been addressed in the design, the architecture, code, tests, and documentation. While the literature describes TIMs (e.g., SafeTIM~\cite{nair2014safetim}), it is unclear how they can be adapted to an organisation and if they fit other needs for traceability (such as change impact analysis). In addition, the evolution of artefacts needs to be sufficiently captured in the information in order to track changes in both artefacts and the links and ensure consistency.

\paragraph{``What are critical decisions when defining a TIM?''} The chosen TIM (e.g., SafeTIM) has a huge impact on how the links can be used later on in the project. At this point, there is no method for how to define a TIM to address the traceability needs of an organisation. Traceability needs include the purpose of establishing trace links (e.g., for change impact analysis or for program comprehension), the process steps in which trace links should be established, maintained, and used, and an alignment with the overall process goals. Since such a method is missing, there is no clear understanding for which decisions are critical when defining a TIM and which impact these decisions will have.  This makes it difficult to foresee how well a TIM will be able to support the organisation in the future. Taken together with the high cost and effort of evolving the TIM, this makes organisations reluctant to commit to a specific TIM.

\paragraph{``Trace between safety analysis artefacts on the same level of abstraction.''}%Item definition contains requirements, but not safety requirements? (No high-level safety requirements? Only in the decomposition?)''
%\rebekka{The item definition is the input to the HARA (Hazard Analysis and Risk Assessment), which is used to understand hazards and derive safety goals. In the item definition, people define what function with use cases and functional requirements they should focus on. The safety goals (output of the HARA) are then typically an input to other safety analyses (like fault trees) to derive safety requirements. All of these things are done on a function/feature level, so typically very early, and for each function/feature. While theoretically one could reuse safety goals or requirements, that is not often done (and there is not a lot of support for traceability between information connected to different functions/features).}
%\rebekka{change the title to ``Trace between safety analysis artifacts on the same level of abstraction'' or something like that?} 
The item definition according to ISO 26262~\cite{ISO26262}, the standard for functional safety in the automotive domain, focuses on single vehicle functions with selected use cases and functional requirements.
The hazard analysis, i.e., the activity in which the top-level safety requirements are defined, is based on this functional description.
However, the high-level functional requirements and their related high-level safety requirements are defined as siblings in a hierarchy of requirements, without explicit trace links between them.
In practice, however, safety-oriented concerns of different functions are related to each other and safety goals and requirements can impact the development of multiple functions.
The lack of traceability makes it difficult to evolve these aspects together.
In order to more easily create and maintain trace links between safety-related information, the functional description should ideally be expressed in a formalised way and tooling and concepts should allow relating cross-cutting aspects at any time during the process.
%\todo[inline]{Even though updated on Joakim's feedback, this still needs some more work.}

\paragraph{``Trace to review status, changes, and decisions.''} It is important to know whether reviews have been passed to understand the current state of the system . Similar to tracing to test results, this enables  engineers to see which aspects of the overall safety argument for the system are covered and what is left to do. Relevant changes that have an impact on these reviews and their status must also be traced in order to understand when a review needs to be repeated. At the same time, important decisions that impact safety need to be traceable, e.g., as design rationales, to help engineers understand why the arguments were constructed the way they are and how the underlying architecture impacts this argument.

\paragraph{``Creating, storing, and accessing baselines.''} A baseline is a snapshot of all ar\-te\-facts relevant at a specific point in time in the development process. Having many different, interrelated artefacts with different lifecycles, worked on by different teams at different locations, and stored in different systems~\cite{knauss2016continuous} makes it difficult to define and store a consistent snapshot and make it available, e.g., to auditors.

\subsection{Continuous Compliance}

Keeping safety-related artefacts up-to-date in a scaled agile setting requires incremental safety analysis that spans all required product variants. There is also a push towards certifying the manufacturer instead of the product itself. The focus group revealed the following challenges:

\paragraph{``Support delta analysis.''} Changes in the system should not necessitate a complete reconstruction of the safety case. Instead, only the relevant parts should be reassessed and the current safety case should only be updated to the extent necessary. Consistent traceability is one cornerstone to solve this issue since it allows to include safety cases in change impact analysis. On the other hand, techniques used in safety analysis, such as formal hazard analysis techniques, should to be able to handle incremental changes. 

\paragraph{``Update safety case on demand.''} As stated above, the goal is to produce safety arguments about deltas and thus only focus on changes and their impacts. This would mean that safety analysis is near-continuous, triggered predominantly by changes. However, even if the technical challenges of delta-analysis are solved, it is currently unclear how often to create a new safety argument. Does the creation of a new or updated safety case depend on the ``size'' of the delta? What is the right balance between a desire for continual safety and the resources needed to produce frequent safety arguments?  How often is often enough?  %JH could be merged, the big question is when to peform the delta analysis, how much has to change.  You don't want to actually be continuous, but near continuous, daily?  Weekly?  When change is > X?  

\paragraph{``Safety case must cover variants.''} Since automotive companies usually work with software product lines, a safety case must cover all relevant variants of a system. That means that regardless how the final system is assembled from different re-usable assets, the safety case must hold. In practice, however, many feature combinations are not relevant. Developers are not always aware of which combinations are relevant, though, and sampling strategies are often unsystematic~\cite{mukelabai2018tackling}. For continuous compliance, safety cases must at least cover those variants that are used in production and must show systematically that these variants are safe.

\paragraph{``Facilitate pre-certification.''} In the medical device domain, where standard bodies govern and enforce the use of safety standards, such bodies have begun to allow \emph{pre-certification}, i.e., a certification of the organisation and their development and quality assurance practices rather than the individual software in an attempt to reduce the time to market.\footnote{See, e.g., \url{https://www.fda.gov/MedicalDevices/DigitalHealth/DigitalHealthPreCertProgram/ucm584020.htm}} Such approaches help to avoid the ``big bang'', all-at-once certification process before releasing products, moving certification steps earlier in the development lifecycle. They also make it easier to push updates to existing software to the customer continuously.

\subsection{(Organisational) Flexibility -- Safe Ecosystem}
\label{sec:open-challenges:safe-ecosystem}

One part of organisational flexibility for automotive OEMs is the ability to use components from suppliers with as little effort as possible in a safe ecosystem. Our participants identified the following challenges:

\paragraph{``Passing safety requirements to suppliers.''} The communication between an OEM and a supplier about requirements for components at the moment is based on the exchange of documents that contain both functional and safety requirements. The supplier transfers these requirements into a requirements management tool and starts using them in the development and the construction of the safety case. However, it is not uncommon that the OEM changes functional and safety requirements. In that case, the supplier receives a new document and has to manually update the requirements database, update the trace links, and understand the impact on the current design~\cite{maro2018software}. Clearly defined software interfaces for the exchange of requirements would improve such updates. A common exchange standard, e.g., similar to the ReqIF format~\cite{ebert2012reqif}, could be a first step. A system that also supports versioning and diffing of such requirements would further reduce the effort required for suppliers.  
%\todo[inline]{This relates strongly to the point later about indicating what functionality is and is not safety-critical.}

\paragraph{``Treat components as safety blackboxes.''}  At the moment, safety-critical components that an OEM buys from a supplier need to be fully transparent in terms of design, safety requirements, and safety cases in order to be integrated into the safety argument for the overall system. Such components can thus not be treated as black boxes and the OEM has to invest considerable effort to integrate the relevant artefacts. In the future, it is desirable that individual components have a clearly defined \emph{safety contract}~\cite{gallina2015using}, e.g., based on assumptions and guarantees, that can be used to seamlessly integrate a component into the safety argument of the overall system. While previous work on this topic exists, a standard for the exchange, verification, and use of such contracts has yet to emerge.

\subsection{(Organisational) Flexibility -- Change Management}

Part of an agile way of working is the ability to react to changes quickly and to adapt what is being built within a short period of time. This requires the ability to also adapt the safety case as needed. The focus group mentioned two challenges in this area:

\paragraph{``Support local decisions and changes.''} Since safety is an overarching concern, decisions about safety and the construction of the safety case are often centralised, e.g., in an architecture runway team. This limits the flexibility of the individual teams to make decisions about implementation details and creates bottle necks in the certification process. Instead, local design decisions and changes should be supported when making the safety case, e.g., by modularising it and giving the individual teams the opportunity to update the safety case locally while maintaining global consistency. %\jp{modularisation of the safety case}

\paragraph{``How to decide which changes need a change request?''} A change request is a formal way to control the change process for product changes that have an impact on other development teams, downstream artefacts and, in particular, the safety case. Since change requests require certain steps to be completed and a high level of rigour to be applied, they are costly and should only be used if necessary. However, this is difficult to determine for any given product change. Organisations therefore err on the side of caution, producing more change requests than necessary. If living traceability is established, however, it should be possible for tools to provide decision support to semi-automatically identify the impact of a change on other teams and the safety case and thus reduce the number of change requests and, consequently, development cost and time.

\subsection{(Organisational) Flexibility -- Way of Working}

As a final building block towards flexibility, the way of working needs to address a number of aspects on a fundamental level. Our participants identified the following challenges in this area:

	\paragraph{``Mixing safety-critical components and requirements with less or non-safety crit\-i\-cal requirements.''} It is common in the automotive domain that components provide safety-critical functionality as well as functionality that is not safety-critical. At the moment, these functionalities are treated differently based on their Automotive Safety Integrity Level (ASIL) as defined in ISO~26262. Functionality assigned ``QM'' is not safety-critical, ASIL A or B is safety-critical and required some validation while ASIL C or D are highly safety-critical and require rigorous validation. Ideally components with different ASIL  should be isolated via the architecture, but this can not always be guaranteed. Requirements management tools and practices currently do not allow a fine-grained assignment of ASIL to components, e.g., on the level of features or even code blocks. This in turn means that all functionality of a component is treated with the same rigour, potentially using resources that could be applied elsewhere.
	 As mentioned in Section~\ref{sec:agile-approaches}, mixed criticality is not directly supported by \rscrum{} or \safescrum{}.
	
	\paragraph{``Reuse of safety requirements and arguments.''} In many cases, safety requirements can be reused across components and functionality. Since safety requirements are often linked to functional requirements, a reuse of the functional requirement can lead to a reuse of the safety requirement.  % (reuse, but also problem vs. solution space, both only defined during impl) \jp{does someone remember what the ``problem vs. solution space'' was about?}  %JH no.  I guess this goes with requirements reuse, if you reuse requirements, you also reuse all the safety requirements linked to that requirement.  Could also reuse just the safety requirements.  If so, how much of the safety arugment has to be redone.  Can they reuse safety arguments?
%	\item ``How to reuse functional safety information, arguments, etc.'' 
	Ideally, parts of the safety argument and the information used to construct them should also be reused. This is particularly true for different variants of a system in a software product line that have slight differences in functionality but are structurally and behaviourally similar. However, an understanding of the changes needed to reuse a safety argument in different circumstances as well as tool support to detect inconsistencies in reused safety requirements and safety arguments is required. %JH exactly.  But this also can introduce copy/paste errors, need some smart tools to indicate which parts should change. 
%	\paragraph{``Validate safety on a simulated model including hardware models''} \jp{Related to \cite{agren2019impediments} according to Eric, but still not sure how to describe this properly.}
%	\item Abstraction via models (e.g., CAD), validate safety at this level? Helps deal with hardware? ->Bottom-up vs. Top-down \jp{Since CAD models are purely hardware, I think we need to approach this differently. Maybe this means that we need to also consider the safety argument on the hardware side?} %JH I'll try, if CAD is only for hardware, maybe this doesn't apply:  If safety is validated all the way down to the level of the code, much effort is needed.  Given the presence of commonly used design models, particularly CAD models, it is desirable to validate and argue for safety at this level of abstraction, saving effort. This is particularly the case for components which have strong degree of hardware, or physical design, in addition to software. This may require CAD tools and processes to be themselves certified and tested for downstream safety.  \jenho{If you want to keep this one, I suggest my text, or a variation, else skip.}
	\paragraph{``Coordination and Modularization.''} Development processes in which several hundred developers across different departments and potentially even organisations are involved require a high degree of coordination and modularisation. Since safety is a cross-cutting concern, achieving safety in a complex system composed of several subsystems is challenging. The modularisation and architectural isolation of functionality mentioned above is a first step, but distributing the work required to construct the safety argument is also necessary. A possible solution is a modularized safety argument. That means that a compositional form of discovering and including safety requirements~\cite{cleland2018discovering} as well as constructing the safety argument~\cite{sharvia2011integrated} is required as well as automated ways of checking the argument for consistency (such as extensions of~\cite{antonino2014improving}). Such a modular approach will allow different teams to work on isolated parts of the safety argument and compose the individual parts into one encompassing subsystems and finally the entire system. Such an approach resembles the idea of ``safety blackboxes'' (cf. Section~\ref{sec:open-challenges:safe-ecosystem}) and would also support the reuse of safety requirements and arguments.

\section{Discussion}
\label{sec:discussion}

In this work, we have extracted common principles, practices and limitations from the literature on safety-critical agile methods and compared this to the experiences of three senior experts from industry, two of whom work in the automotive domain.  What we find is an extended and refined list of principles and challenges as well as a number of solution candidates.

\subsection{Challenges}

In answering RQ1, we find that the literature emphasizes traceability, continuous safety, shared responsibility, and ongoing auditor involvement.  Our findings echo the literature emphasis on traceability, but add specific details and challenges with implementing traceability in large-scale safety-focused agile (RQ2). The idea of ``living traceability'' as an ongoing set of activities that are part of the daily work of the developers bears a strong resemblance to \emph{ubiquitous traceability}~\cite{cleland2014software}, an idea championed in the traceability literature. While challenges to traceability in the automotive domain have been described elsewhere (see, e.g., \cite{maro2018software}), this study adds additional challenges on a more technical level, e.g., about creating baselines. These challenges are nonetheless important, since their solution will decide about wide-spread adoption of traceability practices in industry. 

The focus on scaling is missing in the literature on agile processes for safety-critical systems and current approaches that take scaling into account do not cater to the needs of safety-critical systems~\cite{putta2018benefits}. To address safety in an agile way of working, it should be possible to view traceability in both a bottom-up and top-down way, TIMs should be specific for safety concepts and should come with guidance for design, safety-related traceability should extend horizontally across requirements, should include review statuses, and should account for baselines.  

Our findings also confirm and expand on the area of continuous safety.  On this topic, we can also add specific technical challenges (RQ2). In particular, safety should be analyzed on the delta of small changes with guidance provided on the size of such deltas, safety cases should account for software variants, and the possibility of pre-certification should be considered.  This last point bears similarity to the ongoing auditor involvement practice extracted from the literature.  However, in general, the involvement of auditors was not emphasized, as this practice is less relevant in the automotive domain. Our participant from the medical device industry emphasised this aspect, however, as pre-certification and the ongoing involvement of auditors in the development process can be a key contributor to reduce the time to market in this domain.% \jenho{help.  I got from the workshop that there are no ISO auditors, but they probably still have internal audits?}\jp{That's correct.}.

While the literature emphasized shared safety responsibility, our workshop findings placed more emphasis on organizational flexibility, including ecosystems, change management, and ways of working. This is a direct result of the more complex organisational structure present in the scaled agile environments our practitioners work in. Again, these findings were broken down into more specific challenges (RQ2). In the area of ecosystems, identified challenges centred on passing safety information to suppliers, and receiving safety information from components in a clear and easily understandable format.  

From the change management perspective, decision making should be local when possible, and decision making concerning invoking change requests should be better supported.  The former point most closely echoes the shared responsibility emphasis from the literature, but puts it into the context of a hierarchical organisation in which responsibilities need to be distributed to development teams and some decisions, e.g., about architecture, are made on a product-level~\cite{eckstein2014architecture}.

Practical challenges related to ways of working include dealing with a mix of safety- and non-safety-critical components, reuse of safety requirements and arguments, how to the level of abstraction of safety arguments, and how to manage coordination and modularization.  These challenges are at a level of specificity not found in the current literature.% \jenho{I say this having not read it.  Please correct me if I'm wrong.}

\subsection{Possible Solutions}

Before consulting with our industrial partners, we noted that the literature on safety-focused agile does not consider scale, mixed criticality, or automation.  Our industrial challenges confirmed the first two observations.  Although  many of our identified challenges can lead to automation or benefit from it, this was not identified as a direct challenge in practice. This might be due to the fact that an \emph{increased degree of automation} might be seen as one of the solutions for these challenges. A tool-chain that supports living traceability, helps identify if a safety case needs to be changed, and integrates variants into the handling of safety arguments would be highly beneficial. 

Another possible solution, in particular to the challenges associated with continuous compliance, are techniques that allow the \emph{incremental update of the safety case}. Industry needs the ability to update small parts of the safety case based on individual change requests to reduce the cost and time required for changes and to allow integrating components from suppliers into the system The need for such techniques has also been acknowledged in the defense industry ``as  a  means  of  reducing  the  impact  and  hence  cost  of  re-certification  of  changes  to  systems''~\cite{fenn2007who}. While some work on incremental safety assessment exists, it is either focused on describing formal refinement relations~\cite{lisagor2010incremental} or make an argument for first modularising the safety case~\cite{wilson2009incremental} before taking further steps in this direction. The variability inherent to complex product lines, e.g., in the automotive industry also needs to be taken into account.

The \emph{modularisation of the safety case} also came up as a crucial building block to address the challenges in our data. While a number of solutions have been proposed for modular safety cases (see, e.g., \cite{althammer2008modular,zimmer2011vertical,denney2015towards}), they are not used in practice by our participants, presumably since they are tightly coupled to an underlying architecture~\cite{althammer2008modular} or prescribe a specific notation and toolset~\cite{zimmer2011vertical,denney2015towards}. None of the approaches addresses the needs of a complex product line. Our industrial partners require more generic guidelines that they can adapt to their existing processes, architecture, and tool-chain instead.

\section{Conclusion}

In this paper, we summarise our findings of challenges of applying agile methods to the development of safety-critical systems in large-scale industrial settings. Based on a focus group, we identify a number of challenges in three areas and compare them with what is known from the literature, in particular to \rscrum{} and \safescrum.

We have noted the lack of work combining large-scale agile practices with safety-critical agile practices, even though such a combination is  currently required in many automotive organizations.  Overall, our findings summarize the limitations with current safe agile practices, and list practical, grounded challenges of using such methods in a large-scale context, with a focus on the automotive domain. These challenges can be a foundation for future work and for combining the rigorous approach to safety analysis and verification of \rscrum{} and \safescrum{} with the scaled agile practices of SAFe and LeSS. In particular, a better understanding of establishing traceability throughout the development lifecycle is needed.  In addition, the ability to proof continuous compliance based on updates of safety cases is a necessity. Finally, safe ecosystems, an integrated change management approach, and a way of working based on reuse, coordination, and modularisation will ensure organisational flexibility.

We also identify candidates for solution approaches and point out related work in the area.
The exploration of traceability information models for safety-critical applications is one such starting point, but currently suffers from limited guidance on how to apply this in a practical setting.
The ability to incrementally update the safety cases based on small changes and the incorporation of variability in the safety analysis are further pre-requisites to achieve organisational flexibility.
Furthermore, a modularisation of the safety case would help organisations in including externally sourced components and build incremental safety arguments based on small change requests.

We hope that our findings spur future work in expanding and refining agile methods: to support developing safety critical products at scale, and to consider further challenges as reported by our industrial partners. 

\section*{\ackname}

We thank all participants in our focus group for their insights and their engagement. This work was supported by Software Center (www.software-center.se).

\bibliographystyle{splncs04}
\bibliography{scaled-safe-agile}

\begin{thebibliography}{10}
\providecommand{\url}[1]{\texttt{#1}}
\providecommand{\urlprefix}{URL }
\providecommand{\doi}[1]{https://doi.org/#1}

\bibitem{althammer2008modular}
{Althammer}, E., {Schoitsch}, E., {Sonneck}, G., {Eriksson}, H., {Vinter}, J.:
  Modular certification support — the decos concept of generic safety cases.
  In: 6th IEEE International Conference on Industrial Informatics. pp. 258--263
  (July 2008). \doi{10.1109/INDIN.2008.4618105}

\bibitem{antonino2014improving}
Antonino, P.O., Trapp, M.: Improving consistency checks between safety concepts
  and view based. architecture design. PSAM12, Honolulu, Hawaii, USA
  \textbf{282} (2014)

\bibitem{Broy2007}
Broy, M., Kr{\"{u}}ger, I.H., Pretschner, A., Salzmann, C.: Engineering
  automotive software. Proceedings of the IEEE  \textbf{95}(2),  356--373 (Feb
  2007)

\bibitem{cleland2014software}
Cleland-Huang, J., Gotel, O.C., Huffman~Hayes, J., M{\"a}der, P., Zisman, A.:
  Software traceability: trends and future directions. In: Proceedings of the
  on Future of Software Engineering. pp. 55--69. ACM (2014)

\bibitem{cleland2018discovering}
{Cleland-Huang}, J., {Vierhauser}, M.: Discovering, analyzing, and managing
  safety stories in agile projects. In: IEEE 26th International Requirements
  Engineering Conference (RE). pp. 262--273 (Aug 2018).
  \doi{10.1109/RE.2018.00034}

\bibitem{denney2015towards}
Denney, E., Pai, G.: Towards a formal basis for modular safety cases. In:
  Computer Safety, Reliability, and Security. pp. 328--343. Springer, Cham
  (2015)

\bibitem{Ebert2017}
Ebert, C., Favaro, J.: Automotive software. IEEE Software  \textbf{34}(3),
  33--39 (May 2017). \doi{10.1109/MS.2017.82}

\bibitem{ebert2012reqif}
Ebert, C., Jastram, M.: Reqif: Seamless requirements interchange format between
  business partners. {IEEE Software}  \textbf{29}(5),  82--87 (2012)

\bibitem{eckstein2014architecture}
Eckstein, J.: Architecture in large scale agile development. In: Agile Methods.
  Large-Scale Development, Refactoring, Testing, and Estimation. pp. 21--29.
  Springer, Cham (2014)

\bibitem{fenn2007who}
Fenn, J.L., Hawkins, R., Williams, P., Kelly, T., Banner, M., Oakshott, Y.: The
  who, where, how, why and when of modular and incremental certification. IET
  Conference Proceedings pp. 135--140(5) (January 2007)

\bibitem{rscrum}
Fitzgerald, B., Stol, K.J., O'Sullivan, R., O'Brien, D.: Scaling agile methods
  to regulated environments: An industry case study. In: Int.\ Conf.\ on
  Software Engineering. pp. 863--872. ICSE '13, IEEE Press, Piscataway, NJ, USA
  (2013)

\bibitem{gallina2015using}
Gallina, B., Carlson, J., Hansson, H., et~al.: Using safety contracts to guide
  the integration of reusable safety elements within iso 26262. In: 21st
  Pacific Rim Int.\ Symposium on Dependable Computing (PRDC). pp. 129--138.
  IEEE (2015)

\bibitem{hanssen2018safescrum}
Hanssen, G.K., St{\aa}lhane, T., Myklebust, T.:
  SafeScrum{\textregistered}-Agile Development of Safety-Critical Software.
  Springer (2018)

\bibitem{ISO26262}
{International Organization for Standardization}: Road vehicles -- functional
  safety. ISO26262:2011  (Nov 2011)

\bibitem{kasauli2018safety}
{Kasauli}, R., {Knauss}, E., {Kanagwa}, B., {Nilsson}, A., {Calikli}, G.:
  Safety-critical systems and agile development: A mapping study. In: 2018 44th
  Euromicro Conference on Software Engineering and Advanced Applications
  (SEAA). pp. 470--477 (Aug 2018)

\bibitem{knauss2016continuous}
Knauss, E., Pelliccione, P., Heldal, R., {\AA}gren, M., Hellman, S., Maniette,
  D.: Continuous integration beyond the team: a tooling perspective on
  challenges in the automotive industry. In: 10th ACM/IEEE Int.\ Symp.\ on
  Empirical Software Engineering and Measurement. p.~43. ACM (2016)

\bibitem{lisagor2010incremental}
Lisagor, O., Bozzano, M., Bretschneider, M., Kelly, T.: Incremental safety
  assessment: Enabling the comparison of safety analysis results. In: 28th
  International System Safety Conference (ISSC) (2010)

\bibitem{maro2018software}
Maro, S., Stegh{\"o}fer, J.P., Staron, M.: Software traceability in the
  automotive domain: challenges and solutions. JSS  \textbf{141},  85--110
  (2018)

\bibitem{mukelabai2018tackling}
Mukelabai, M., Ne{\v{s}}ic, D., Maro, S., Berger, T., Stegh{\"o}fer, J.P.:
  Tackling combinatorial explosion: a study of industrial needs and practices
  for analyzing highly configurable systems. In: 33rd IEEE/ACM Int.\ Conf.\ on
  Automated Software Engineering (ASE) (2018)

\bibitem{myklebust2016agile}
Myklebust, T., St{\aa}lhane, T., Lyngby, N.: An agile development process for
  petrochemical safety conformant software. In: 2016 Annual Reliability and
  Maintainability Symposium (RAMS). pp.~1--6. IEEE (2016)

\bibitem{nair2014safetim}
Nair, S., de~la Vara, J.L., Melzi, A., Tagliaferri, G., de-la Beaujardiere, L.,
  Belmonte, F.: Safety evidence traceability: Problem analysis and model. In:
  Requirements Engineering: Foundation for Software Quality. pp. 309--324.
  Springer, Cham (2014)

\bibitem{putta2018benefits}
Putta, A., Paasivaara, M., Lassenius, C.: Benefits and challenges of adopting
  the scaled agile framework (safe): Preliminary results from a multivocal
  literature review. In: Product-Focused Software Process Improvement. pp.
  334--351. Springer, Cham (2018)

\bibitem{schoonderwoert2018agile}
Schooenderwoert, N.V., Shoemaker, B.: Agile Methods for Safety-Critical
  Systems: A Primer Using Medical Device Examples. CreateSpace Independent
  Publishing Platform (2018)

\bibitem{sharvia2011integrated}
Sharvia, S., Papadopoulos, Y.: Integrated application of compositional and
  behavioural safety analysis. In: Dependable Computer Systems, pp. 179--192.
  Springer (2011)

\bibitem{staalhane2012application}
St{\aa}lhane, T., Myklebust, T., Hanssen, G.: The application of safe scrum to
  iec 61508 certifiable software. In: 11th International Probabilistic Safety
  Assessment and Management Conference and the Annual European Safety and
  Reliability Conference. pp. 6052--6061 (2012)

\bibitem{wilson2009incremental}
{Wilson}, A., {Preyssler}, T.: Incremental certification and integrated modular
  avionics. IEEE Aerospace and Electronic Systems Magazine  \textbf{24}(11),
  10--15 (Nov 2009)

\bibitem{zimmer2011vertical}
Zimmer, B., B{\"u}rklen, S., Knoop, M., H{\"o}fflinger, J., Trapp, M.: Vertical
  safety interfaces -- improving the efficiency of modular certification. In:
  Computer Safety, Reliability, and Security. pp. 29--42. Springer, Berlin,
  Heidelberg (2011)

\end{thebibliography}

\end{document}